\documentclass[12pt]{article}

\usepackage{epsfig}

\usepackage{bbm}
\usepackage{amssymb,amsmath}

\newcommand{\be}{\begin{equation}}
\newcommand{\ee}{\end{equation}}
\newcommand{\ba}{\begin{eqnarray}}
\newcommand{\ea}{\end{eqnarray}}
\newcommand{\no}{\nonumber\\}

\newcommand{\mnu}{\mathcal{M}_\nu}
\newcommand{\dsol}{\Delta m^2_\odot}
\newcommand{\datm}{\Delta m^2_\mathrm{atm}}

\begin{document}

\title{\normalsize \hfill UWThPh-2006-28 \\[8mm]
\LARGE A three-parameter model for the neutrino mass matrix}

\author{
W.~Grimus$^{(1)}$\thanks{E-mail: walter.grimus@univie.ac.at}
\ and
L.~Lavoura$^{(2)}$\thanks{E-mail: balio@cftp.ist.utl.pt}
\\*[4mm]
$^{(1)}$ \small
Fakult\"at f\"ur Physik, Universit\"at Wien \\
\small
Boltzmanngasse 5, A--1090 Wien, Austria
\\*[3mm]
$^{(2)}$ \small
Universidade T\'ecnica de Lisboa
and Centro de F\'\i sica Te\'orica de Part\'\i culas \\
\small
Instituto Superior T\'ecnico, 1049-001 Lisboa, Portugal
}

\date{24 May 2007}

\maketitle

\begin{abstract}
Using the type-II seesaw mechanism
with three Higgs doublets $\phi_\alpha$
($\alpha = e,\,\mu,\,\tau$)
and four Higgs triplets,
we build a model for lepton mixing
based on a 384-element horizontal symmetry group,
generated by the permutation group $S_3$ and by six 
$\mathbbm{Z}_2$ transformations.
The charged-lepton mass matrix is diagonal;
the symmetries of the model would
require all the three masses $m_\alpha$
to be equal,
but different vacuum expectation values of the $\phi_\alpha$
allow the $m_\alpha$ to split.
The number of parameters
in the Majorana neutrino mass matrix $\mnu$
depends on two options: 
full breaking of the permutation group $S_3$,
or leaving a $\mu$--$\tau$ interchange symmetry intact;
and hard or spontaneous violation of $CP$.
We discuss in detail the case
with the minimal number of three parameters,
wherein $\mnu$ is real,
symmetric under $\mu$--$\tau$ interchange,
and has equal diagonal elements.
In that case,
$CP$ is conserved in lepton mixing,
atmospheric neutrino mixing is maximal,
and $\theta_{13} = 0$;
moreover,
the type of neutrino mass spectrum
and the absolute neutrino mass scale 
are sensitive functions of the solar mixing angle.
\end{abstract}

\newpage

\section{Introduction}

There are two puzzles associated with neutrinos:
why are their masses so much smaller
than those of the charged fermions,
and why does the lepton mixing matrix
feature large mixing angles---for reviews
see~\cite{reviews}---in contrast to the quark mixing matrix.
It is possible that both puzzles
are solved through the same mechanism.
In this paper we envisage the type-II seesaw mechanism
as a possible solution.\footnote{In our model
we do not allow for a type-I seesaw mechanism;
we assume right-handed neutrino singlets not to exist.}
We use horizontal symmetries
to enforce certain features of lepton mixing,
in particular maximal atmospheric neutrino mixing
and $\theta_{13} = 0$.
In order to achieve this,
we enlarge the scalar sector of the Standard Model
by adding to it four Higgs triplets
and by using altogether three Higgs doublets.
Our model has a permutation group $S_3$
together with six cyclic symmetries $\mathbbm{Z}_2$,
which commute with each other but not with $S_3$;
the result is a large discrete symmetry group
with 384 elements.
This setting allows to obtain
four different neutrino mass matrices,
depending on the assumed breaking of the horizontal symmetries
and of the symmetry $CP$.
Amazingly,
by breaking the horizontal symmetries softly
by terms of dimension two,
while leaving a residual $\mu$--$\tau$ interchange symmetry
to be broken at low energy in the charged-lepton sector,
we arrive at a \emph{viable} neutrino mass matrix
with \emph{only three real parameters}.

In section~\ref{II} we make a general discussion
of the type-II seesaw mechanism
with an arbitrary number of Higgs doublets and triplets.
Our model,
with its multiplets,
symmetries,
and Lagrangian is explained in section~\ref{model}.
In section~\ref{three-parameter}
we investigate in detail the most predictive case
of a three-parameter neutrino mass matrix.
A generalization thererof is considred in section~\ref{complex}.
The conclusions are presented in section~\ref{concl}.

\section{The type-II seesaw mechanism}
\label{II}

We first review the type-II seesaw mechanism~\cite{II,ma}
for small neutrino masses.
We assume the existence---in the electroweak theory---of 
several Higgs doublets $\phi_\alpha$ with hypercharge $1/2$,
together with several Higgs triplets $\Delta_i$ with hypercharge $1$.
Let the neutral components of the $\phi_\alpha$
have vacuum expectation values (VEVs) $v_\alpha$
and the neutral components of the $\Delta_i$ have VEVs $\delta_i$.
Just because of the hypercharge symmetry,
the vacuum potential $V_0$ must be of the
form\footnote{We use the summation convention.}
\ba
V_0 &=&
\left( \mu_\phi^2 \right)_{\alpha\beta} v_\alpha^\ast v_\beta
+ \left( \mu_\Delta^2 \right)_{ij} \delta_i^\ast \delta_j
+ \left( t_{i\alpha\beta} \delta_i^\ast v_\alpha v_\beta
+ \mbox{c.c.} \right)
\no & &
+ \lambda_{\alpha\beta\gamma\delta}
v_\alpha^\ast v_\beta v_\gamma^\ast v_\delta
+ \lambda_{ijkl} \delta_i^\ast \delta_j \delta_k^\ast \delta_l
+ \lambda_{\alpha\beta ij} v_\alpha^\ast v_\beta
\delta_i^\ast \delta_j.
\ea
The matrices $\mu_\phi^2$ and $\mu_\Delta^2$ are Hermitian and,
likewise,
the $\lambda$ coefficients must obey various conditions
in order that $V_0$ should be real.
The VEVs of the triplets are determined by 
\be
0 = \frac{\partial V_0}{\partial \delta_i^\ast} =
\left( \mu_\Delta^2 \right)_{ij} \delta_j
+ t_{i\alpha\beta} v_\alpha v_\beta
+ 2 \lambda_{ijkl} \delta_j \delta_k^\ast \delta_l
+ \lambda_{\alpha\beta ij} v_\alpha^\ast v_\beta \delta_j.
\label{stabil}
\ee
Contrary to $\mu_\phi^2$,
we assume the matrix $\mu_\Delta^2$ to be positive definite 
so that,
in the absence of the $t_{i\alpha\beta}$ terms,
the only solution to equations~(\ref{stabil})
would be for all the $\delta_i$ to vanish.
The VEVs $v_\alpha$
are of order of the electroweak scale $v \approx 174\, \mathrm{GeV}$,
or smaller.
Assuming the $t_{i\alpha\beta}$ to be of order $M$
and the eigenvalues of $\mu_\Delta^2$ to be of order $M^2$,
where $M$ is a mass scale much larger than $v$~\cite{ma},
the approximate solution to equations~(\ref{stabil})
is given by~\cite{GL05}
\be
\delta_i \approx - \left( \mu_\Delta^2 \right)^{-1}_{ij}
t_{j\alpha\beta} v_\alpha v_\beta.
\label{sol}
\ee
From equation~(\ref{sol}),
the $\delta_i$ are of order $v^2 / M \ll v$.
If,
furthermore,
all the $\lambda$ coefficients are of order unity or smaller,
then the approximate solution~(\ref{sol})
will be corrected on its right-hand side
only by terms suppressed by a factor $v^2 / M^2 \ll 1$.

Under an $SU(2)$ gauge transformation,
the left-handed lepton doublets $D_{L \alpha}$
transform as $D_{L\alpha} \to W D_{L\alpha}$
while the Higgs triplets transform as
$\Delta_i \to W \Delta_i W^\dagger$,
where $W$ is an $SU(2)$ matrix.
Therefore,
the Higgs triplets have Yukawa couplings
of the form $D_{L\alpha}^T C^{-1} \varepsilon \Delta_i D_{L\beta}$,
where $C$ is the charge-conjugation matrix in Dirac space
and $\varepsilon$ is the $2 \times 2$ antisymmetric matrix
in gauge-$SU(2)$ space.
The VEVs $\delta_i$ being very small,
the above Yukawa couplings generate very small neutrino mass terms
$\delta_i\, \nu_{L\alpha}^T C^{-1} \nu_{L\beta}$,
of order $v^2 / M$ times a typical Yukawa-coupling constant.
The neutrino masses being of order $0.1\, \mathrm{eV}$,
$M$ could easily be of order $10^{14}\, \mathrm{GeV}$~\cite{ma},
thus fully justifying the approximate solution~(\ref{sol}).

\section{The model}
\label{model}

Our model follows closely,
in the symmetries that it utilizes,
a previous model of ours~\cite{HPS-model}.
We have three left-handed lepton doublets $D_{L\alpha}$,
three right-handed charged-lepton singlets $\alpha_R$,
and three Higgs doublets $\phi_\alpha$
($\alpha = e, \mu, \tau$).\footnote{Constraints on multi-Higgs doublet
models from electroweak precision tests are not very stringent: Higgs
bosons with large $ZZ$ couplings must have an average mass in the
range allowed for the mass of the Standard Model Higgs boson~\cite{ewtest}.}
There are four Higgs triplets,
$\Delta_\alpha$ and $\Delta_4$.\footnote{The scalar content
of our model resembles that of the $A_4$ model of~\cite{hirsch}.
However,
in that model,
three gauge triplets are used instead of our $\Delta_4$.}
The symmetries of the model consist of a permutation group $S_3$
acting simultaneously on all indices $\alpha$,
three $\mathbbm{Z}_2$ symmetries
\be\label{z1}
\mathbf{z}_\alpha^{(1)}: \quad
\phi_\alpha \to - \phi_\alpha, \
\alpha_R \to - \alpha_R,
\ee
and another three $\mathbbm{Z}_2$ symmetries
\be\label{z2}
\mathbf{z}_\alpha^{(2)}: \quad
D_{L\alpha} \to - D_{L\alpha}, \
\alpha_R \to - \alpha_R, \ \mathrm{and} \
\Delta_\beta \to - \Delta_\beta \ \mathrm{iff} \ \beta \neq \alpha.
\ee
Notice that $\Delta_4$ is invariant under all these symmetries.
In appendix~A we make a study of the full symmetry group of our model.

The Yukawa Lagrangian invariant under all these symmetries is
\ba
\mathcal{L}_\mathrm{Yukawa} &=&
- y_0 \bar D_{L\alpha} \phi_\alpha \alpha_R
+ \frac{1}{2}\, y_1
D_{L\alpha}^T C^{-1} \varepsilon \Delta_4 D_{L\alpha}
\no & &
+ y_2 \left(
D_{Le}^T C^{-1} \varepsilon \Delta_\mu D_{L\tau}
+ D_{L\mu}^T C^{-1} \varepsilon \Delta_\tau D_{Le}
+ D_{L\tau}^T C^{-1} \varepsilon \Delta_e D_{L\mu}
\right) + \mathrm{H.c.}
\ea
Thus,
the charged-lepton mass matrix is automatically diagonal,
the charged lepton $\alpha$ having mass
$m_\alpha = \left| y_0 v_\alpha \right|$.
On the other hand,
the neutrino mass matrix is
\be
\mathcal{M}_\nu = \left( \begin{array}{ccc}
y_1 \delta_4 & y_2 \delta_\tau & y_2 \delta_\mu \\
y_2 \delta_\tau & y_1 \delta_4 & y_2 \delta_e \\
y_2 \delta_\mu & y_2 \delta_e & y_1 \delta_4
\end{array} \right),
\ee
all its diagonal matrix elements being equal.

Due to the symmetries of our model,
the coupling constants $t_{i\alpha\beta}$ of the previous section
assume the very simple form
\be
t_{i\alpha\beta} = t \delta_{i4} \delta_{\alpha\beta}.
\ee
Hence,
from equation~(\ref{sol}),
\be
\delta_i = - t v_\alpha v_\alpha \left( \mu_\Delta^2 \right)^{-1}_{i4}.
\label{di}
\ee
Ordering the triplet fields as
$\left( \Delta_e, \Delta_\mu, \Delta_\tau, \Delta_4 \right)$,
the symmetries of our model would enforce
\be
\mu_\Delta^2 = \mathrm{diag} 
\left( \mu_1^2,\ \mu_1^2,\ \mu_1^2,\ \mu_2^2 \right),
\ee
which is not satisfactory since it would lead,
through equation~(\ref{di}),
to $\delta_e = \delta_\mu = \delta_\tau = 0$.
We must have $\left( \mu_\Delta^2 \right)^{-1}_{i4} \neq 0$
for $i = e, \mu, \tau$.
In order to solve this problem,
we assume the symmetries of the model to be broken softly,
only by terms of dimension two.
Without any residual symmetry,
this means that both matrices $\mu_\Delta^2$ and $\mu_\phi^2$
become fully general,
while all other couplings remain unchanged.

However,
in order to simplify our model and render it more predictive,
we may assume that the interchange symmetry 
$\mu \leftrightarrow \tau$~\cite{early,GL01,HPS,joshipura},
which is a subgroup of our permutation group $S_3$,
is kept unbroken in $\mu_\Delta^2$ and $\mu_\phi^2$.
Then,
\be
\left( \mu_\Delta^2 \right)^{-1} = \left( \begin{array}{cccc}
a & b & b & c \\
b^\ast & d & e & f \\
b^\ast & e & d & f \\
c^\ast & f^\ast & f^\ast & g
\end{array} \right)
\label{mD}
\ee
($a$,
$d$,
$e$,
and $g$ are real),
so that $\delta_e = - t c v_\alpha v_\alpha$,
$\, \delta_\mu = \delta_\tau = - t f v_\alpha v_\alpha$,
and the neutrino mass matrix is $\mu$--$\tau$ symmetric.
This immediately leads to the predictions
$\theta_{23} = \pi / 4$ and $\theta_{13} = 0$.
The $\mu$--$\tau$ interchange symmetry
is supposed to be spontaneously broken
through the VEVs of the Higgs doublets:
$v_\mu \neq v_\tau$.
The Higgs potential is rich enough to allow for this outcome---in
appendix~B we demonstrate this
by working out a simplified case.
Of course,
the spontaneous breaking at the electroweak scale
of the $\mu$--$\tau$ interchange symmetry will seep,
through radiative corrections,
into the rest of the theory,
so that at loop level the matrix $\left( \mu_\Delta^2 \right)^{-1}$
will not any more be of the form in equation~(\ref{mD}),
and then $\delta_\mu \neq \delta_\tau$.
But,
both because this is a loop effect,
and because it is a correction of order 
of the ratio of the electroweak scale
to the much larger mass terms in $\mu_\Delta^2$,
we may expect $\delta_\mu - \delta_\tau$ to remain negligible.

In a further simplification of our model,
we may also assume $CP$ violation to be spontaneous:
the matrix $\left( \mu_\Delta^2 \right)^{-1}$ is then real,
but the VEVs $v_\alpha$ display non-trivial relative phases.
Then,
$\delta_e$,
$\delta_\mu$,
$\delta_\tau$,
and $\delta_4$ will all have the same phase---the phase
of $v_\alpha v_\alpha$.
That phase may be rephased away from $\mnu$,
so that the neutrino mass matrix becomes real.
Thus,
spontaneous $CP$ breaking in our model
yields the remarkable outcome that,
even though there is $CP$ violation,
it remains absent from the mass matrices
and from lepton mixing.\footnote{This is not an original situation;
in the classical Branco model of $CP$ violation~\cite{branco},
spontaneous $CP$ breaking also does not find a way
into the quark mixing matrix.}

One thus obtains the following four possibilities:
\begin{enumerate}
\item The general case,
in which $CP$ violation is hard
and $\mu$--$\tau$ symmetry
is allowed to be broken in $\mu_\Delta^2$.
Then,
\be
\mnu = \left( \begin{array}{ccc}
m & p e^{i \psi} & q e^{i \chi} \\
p e^{i \psi} & m & r e^{i \rho} \\
q e^{i \chi} & r e^{i \rho} & m
\end{array} \right),
\label{mnu1}
\ee
with real $m$,
$p$,
$q$,
and $r$.
This case should not be very predictive,
since it has seven parameters to predict nine 
observables---three neutrino masses,
three lepton mixing angles,
one CKM-type phase,
and two Majorana phases.
\item The case in which
$\mu$--$\tau$ symmetry is allowed to be broken in $\mu_\Delta^2$
but $CP$ violation is spontaneous.
Then,
$\psi$,
$\chi$,
and $\rho$ in~(\ref{mnu1}) vanish.
There is no $CP$ violation in lepton mixing.
The four parameters $m$,
$p$,
$q$,
and $r$ allow one to predict
six observables---three neutrino masses
and three lepton mixing angles.
\item The case in which $\mu$--$\tau$ interchange symmetry
is preserved in $\mu_\Delta^2$,
while $CP$ violation is allowed to be hard.
Then,
\be
\mnu = \left( \begin{array}{ccc}
x & y & y \\
y & x & w \\
y & w & x
\end{array} \right),
\label{mnu2}
\ee
with complex parameters
$x$,
$y$,
and $w$.
There are in this case five parameters---three moduli and two phases.
\item The most predictive case,
in which $CP$ violation is spontaneous
and $\mu$--$\tau$ interchange symmetry
is preserved down to the electroweak scale.
The neutrino mass matrix is the one in equation~(\ref{mnu2})
but with real $x$,
$y$,
and $w$.
The neutrino mass matrix has only three parameters.
\end{enumerate}

\section{The three-parameter neutrino mass matrix}
\label{three-parameter}

In this section we concentrate on case~4 of the previous section,
i.e.~on the neutrino mass matrix of equation~(\ref{mnu2})
with real $x$,
$y$,
and $w$.
The algebra of the diagonalization
of a general $\mu$--$\tau$ symmetric neutrino mass matrix
has been worked out in~\cite{GL01},
and we only need to adapt it to the simpler case~4.
In the following,
the solar mixing angle---which is defined
to be in the first quadrant---is denoted $\theta$,
the neutrino masses are $m_{1,2,3}$,
the solar mass-squared difference is $\dsol = m_2^2 - m_1^2 > 0$,
and the atmospheric mass-squared difference
is
\be
\datm = \left| m_3^2 - \frac{m_1^2 + m_2^2}{2} \right|
= \epsilon \left( m_3^2 - \frac{m_1^2 + m_2^2}{2} \right),
\label{da}
\ee
where $\epsilon = +1$ indicates a normal neutrino mass ordering
and $\epsilon = -1$ an inverted ordering.

Equations~(3.9)--(3.11) and~(3.15) of~\cite{GL01} yield,
respectively,
\ba
m_3 &=& \left| x - w \right|,
\label{m3}
\\
m^2_{1,2} &=& \frac{x^2 + 4 y^2 + \left( x + w \right)^2 \mp \dsol}{2},
\label{m12}
\\
\tan{2 \theta} &=& 
\frac{2 \sqrt{2} \left| y \right|}{w}\
\mathrm{sign} \left( 2x + w \right),
\label{theta}
\\
\dsol \cos{2 \theta} &=& w \left( 2 x + w \right).
\label{dsol}
\ea
Experimentally we know that $\theta$ is in the first octant.
Hence $\tan{2 \theta} > 0$
and equations~(\ref{theta}) and~(\ref{dsol}) both give
\be
\mbox{sign} \left( 2x + w \right) = \mbox{sign}\, w.
\ee
Again from equations~(\ref{theta}) and~(\ref{dsol}),
\ba
\left| y \right| &=& \frac{|w| \tan{2 \theta}}{2\sqrt{2}},
\label{y} \\
x &=& \frac{\dsol \cos 2\theta - w^2}{2w}.
\label{x}
\ea
From equations~(\ref{da}),
(\ref{m3}),
and~(\ref{m12}),
we find
\be
\epsilon \datm = \frac{w^2}{2} - 3 x w - 2 y^2.
\label{datm}
\ee
Inserting equations~(\ref{y}) and~(\ref{x})
into equation~(\ref{datm}),
we obtain the value of $w$:
\be
\label{w}
w^2 = \frac{4 \epsilon \datm + 6 \dsol \cos{2 \theta}}
{8 - \tan^2{2 \theta}}.
\ee
Since $\datm \gg \dsol$,
the numerator of equation~(\ref{w}) has the sign of $\epsilon$;
hence its denominator must also have the sign of $\epsilon$.
That denominator vanishes when $\sin^2 \theta = 1/3$,
i.e.~when $\theta$ is just
the Harrison--Perkins--Scott (HPS) solar mixing angle~\cite{HPS}. 
We thus conclude that,
in our model,
\begin{itemize}
\item
if the neutrino mass spectrum is normal,
then the solar mixing angle is smaller than its HPS value;
\item
if the neutrino mass spectrum is inverted,
then $\sin^2{\theta} > 1/3$.
\end{itemize}
This remarkable result relates 
the type of neutrino mass spectrum
to the value of the solar mixing angle.

From equations~(\ref{m3}),
(\ref{m12}),
(\ref{y}),
and~(\ref{x}),
\be
m_{1,2} = \frac{1}{2} \left(
\frac{|w|}{\cos{2 \theta}}
\mp \frac{\dsol \cos{2 \theta}}{|w|}
\right)
\quad \mbox{and} \quad
m_3 = \frac{3 |w|}{2} - \frac{\dsol \cos{2 \theta}}{2 |w|}.
\ee
This gives the overall scale of the neutrino masses.
Since $|w|$ diverges when $\tan^2{2 \theta} \to 8$,
we see that \emph{in our model the neutrino mass spectrum
becomes quasi-degenerate when the solar mixing angle
approaches its Harrison--Perkins--Scott value}.

One easily sees the reason why our model
displays a singularity when $\sin^2{\theta} = 1/3$.
The most general neutrino mass matrix
leading to HPS lepton mixing is
\be
\mnu = \left( \begin{array}{ccc}
x & y & y \\
y & x+u & y-u \\
y & y-u & x+u
\end{array} \right).
\label{HPS}
\ee
Our $\mnu$ in equation~(\ref{mnu2})
has equal diagonal matrix elements.
Hence,
if it were to accept $\sin^2{\theta} = 1/3$,
it would have to correspond to $u = 0$ in equation~(\ref{HPS}).
But the $\mnu$ of equation~(\ref{HPS}) with $u = 0$
leads to two equal neutrino masses,
hence it is unrealistic.
There is therefore a contradiction with experiment
in the assumption that our $\mnu$ of equation~(\ref{mnu2})
might be compatible with HPS mixing.

Experimentally $\sin^2{\theta}$ is close to $1/3$,
therefore there is the danger that
our neutrino masses are too large
and saturate the cosmological bound~\cite{raffelt}.
As a numerical exercise,
we take the $1\,\sigma$ bound on solar mixing from~\cite{schwetz}:
\be
\label{numerical}
0.27 < \sin^2{\theta} < 0.32 \; \Leftrightarrow
3.73 < \tan^2{2 \theta} < 6.72.
\ee
The mean value of $\theta$ is given by $\sin^2{\theta} = 0.30$ 
and $\tan^2{2 \theta} = 5.25$.
Note that the upper $2\,\sigma$ limit
$\sin^2{\theta} = 0.36$ gives $\tan^2{2 \theta} = 11.76$,
which is already significantly larger than 8.
Thus,
there is experimentally ample room
for the neutrino masses to be sufficiently small.
This happens because $\tan^2{2 \theta}$
is a rapidly varying function of $\theta$.

\begin{figure}[t]
\centering
\epsfig{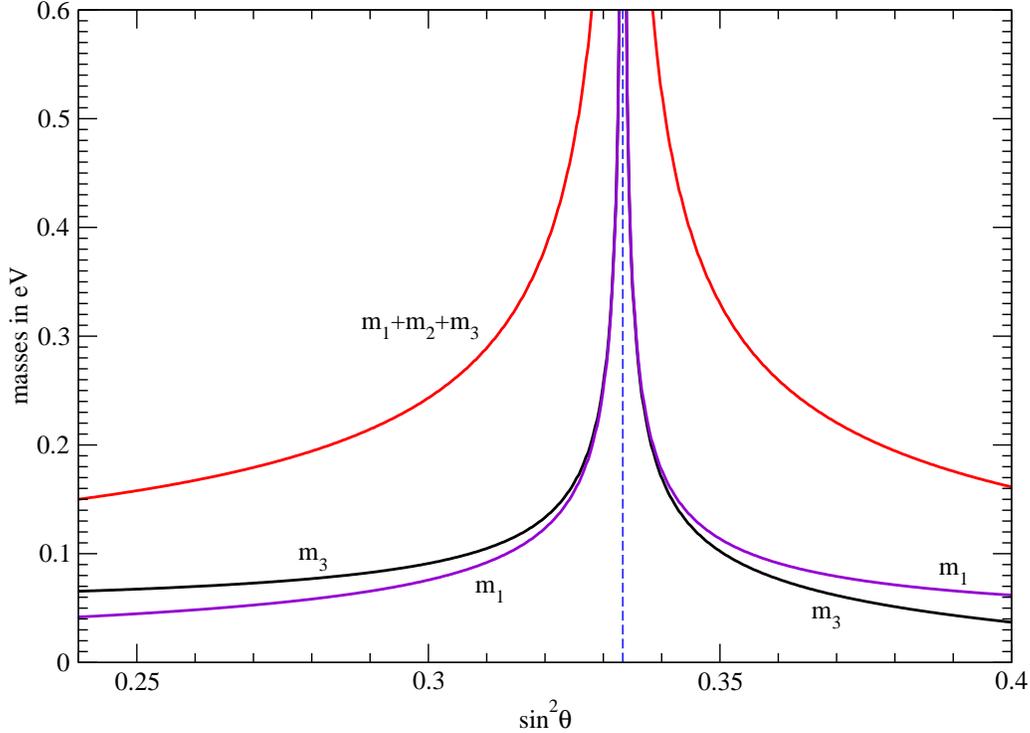}

\vspace{4mm}

\caption{Plot of $m_1$,
$m_3$,
and $\sum_i m_i$
as a function of $\sin^2{\theta}$ in our three-parameter model.
The values of $\datm$ and $\dsol$
have been fixed at $2.5 \times 10^{-3}$ and $7.9 \times 10^{-5}$,
respectively,
in eV$^2$.
The dashed vertical line
indicates the singularity at $\sin^2{\theta} = 1/3$.}
\label{fig1}
\end{figure}

In figure~\ref{fig1} we have plotted $m_1$,
$m_3$,
and $m_1 + m_2 + m_3$ against $\sin^2{\theta}$ in our model.
We have used the best-fit values
$\datm = 2.5 \times 10^{-3}\, \mathrm{eV}^2$
and $\dsol = 7.9 \times 10^{-5}\, \mathrm{eV}^2$
from~\cite{schwetz};
for $\theta$ we have used the $3\, \sigma$ bounds
$0.24 < \sin^2{\theta} < 0.40$,
from the same source.\footnote{A different model in which the neutrino mass
spectrum is normal or inverted
depending on whether $\sin^2{\theta}$ is smaller or larger than $1/3$,
and the neutrinos become degenerate in the limit
$\sin^2{\theta} \to 1/3$,
has been suggested in~\cite{frigerio}.}

Another important observable is $m_{\beta\beta}$,
the effective mass relevant for neutrinoless $2\beta$ decay.
This is equal to the modulus
of the $\left( e,e \right)$ matrix element of $\mnu$,
i.e.,
in our case,
to $|x|$.
Thus,
\be
m_{\beta\beta} = \frac{|w|}{2} - \frac{\dsol \cos{2 \theta}}{2 |w|}.
\ee
Since
\be
|w| \approx 2\, \sqrt{\frac{\datm}{\left| 8 - \tan^2{2 \theta} \right|}}
\gg \sqrt{\dsol \cos{2 \theta}},
\ee
we see that in our model we have the relation
\be
m_{\beta\beta} \approx m_3 / 3.
\ee
This same relation has recently been obtained
in a different model~\cite{helmut}.

One may ask oneself whether the neutrino mass matrix displays
any characteristic texture in the limit $\tan^2{ 2 \theta} \to 8$.
A glance at equations~(\ref{y}), (\ref{x}), and~(\ref{w}) 
allows one to conclude that,
in that limit,
all the matrix elements of $\mnu$ diverge.
Moreover, from equations~(\ref{x}) and~(\ref{y}),
respectively, we obtain
\be
\frac{x}{w} \to - \frac{1}{2},
\quad
\frac{|y|}{|w|} \to 1
\ee
for $\tan^2{ 2 \theta} \to 8$, 
from where the texture of $\mnu$ in that limit can be read off.

\section{Extension to the complex case}
\label{complex}

In this section we investigate what happens when
one allows the neutrino mass matrix of equation~(\ref{mnu2})
to have complex matrix elements.
Does the intriguing feature
of quasi-degenerate neutrinos in the limit of HPS mixing,
found in the previous section for the case of real matrix elements,
still hold true?
We shall see that it does not;
indeed,
the general neutrino mass matrix~(\ref{mnu2})
does not seem to have much predictive power
beyond $U_{e3} = 0$ and maximal atmospheric neutrino mixing.

The symmetric matrix
\be
\mnu = \left( \begin{array}{ccc}
x & y & y \\ y & z & w \\ y & w & z
\end{array} \right)
\label{mnu}
\ee
is diagonalized in the following way:
\be
U^T \mnu\, U = \mbox{diag} \left( m_1, m_2, m_3 \right),
\label{diag}
\ee
the matrix $U$ being unitary
while the $m_j$ ($j = 1, 2, 3$) are real and non-negative.
Due to the special form of $\mnu$,
wherein $\left( \mnu \right)_{12} = \left( \mnu \right)_{13}$
and $\left( \mnu \right)_{22} = \left( \mnu \right)_{33}$,
$U$ is of the form
\be
U =
\mbox{diag} \left( e^{i \varphi}, e^{i \vartheta}, e^{i \vartheta} \right)
\left( \begin{array}{ccc}
c & s & 0 \\ - r s & r c & r \\ - r s & r c & - r
\end{array} \right) 
\mbox{diag} \left( e^{i \Sigma_1}, e^{i \Sigma_2}, e^{i \Sigma_3} \right),
\label{u}
\ee
where $c = \cos{\theta}$,
$s = \sin{\theta}$,
and $r = 2^{-1/2}$.
From equations~(\ref{mnu})--(\ref{u}), we find
\ba
x &=&
e^{- 2 i \varphi} \left( c^2 m_1  e^{- 2 i \Sigma_1}
+ s^2 m_2 e^{- 2 i \Sigma_2} \right),
\\
z &=& \frac{e^{- 2 i \vartheta}}{2} \left(
s^2 m_1 e^{- 2 i \Sigma_1}
+ c^2 m_2 e^{- 2 i \Sigma_2}
+ m_3 e^{- 2 i \Sigma_3} \right).
\ea
We define $\bar m_j \equiv m_j e^{- 2 i \Sigma_j}$ for $j = 1, 2, 3$.
We also define $\chi \equiv 2 \left( \vartheta - \varphi \right)$.
Then,
the condition $x = z$,
which makes the $\mnu$ of equation~(\ref{mnu})
identical with the one of equation~(\ref{mnu2}),
is equivalent to
\be
\bar m_1 \left( 2 c^2 e^{i \chi} - s^2 \right)
+ \bar m_2 \left( 2 s^2 e^{i \chi} - c^2 \right)
- \bar m_3 = 0.
\label{cd}
\ee
Thus,
the condition $x = z$ is equivalent to the existence of four phases $\chi$
and $\Sigma_{1,2,3}$ such that the condition~(\ref{cd}) is satisfied.
That condition states that
it is possible to draw a triangle in the complex plane,
the sides of that triangle having lengths $\sqrt{A} m_1$,
$\sqrt{B} m_2$,
and $m_3$,
where
\be
\begin{array}{rcl}
A &=& 4 c^4 + s^4 - 4 c^2 s^2 \cos{\chi},
\\*[1mm]
B &=& 4 s^4 + c^4 - 4 c^2 s^2 \cos{\chi}.
\end{array}
\label{AB}
\ee
Therefore,
one may eliminate the phases $\Sigma_{1,2,3}$
from condition~(\ref{cd}) by writing
the sole ``triangle inequality''~\cite{book}
\be
A^2 m_1^4 + B^2 m_2^4 + m_3^4 - 2 \left( A B m_1^2 m_2^2
+ A m_1^2 m_3^2 + B m_2^2 m_3^2 \right) \le 0.
\label{ineq1}
\ee
Using
\be
\begin{array}{rcl}
m_1^2 &=& m_3^2 - \epsilon \datm - \frac{1}{2}\, \dsol,
\\*[2mm]
m_2^2 &=& m_3^2 - \epsilon \datm + \frac{1}{2}\, \dsol,
\end{array}
\ee
the inequality~(\ref{ineq1}) takes the form
\be
k_4 m_3^4 + 2 k_2 m_3^2 + k_0 \le 0,
\label{in2}
\ee
where
\ba
k_4 &=& 1 - 2 \left( A + B \right) + \left( A - B \right)^2,
\\
k_2 &=& \left[ A + B - \left( A - B \right)^2 \right] \epsilon \datm
+ \frac{1}{2} \left( A - B \right)
\left( 1 - A - B \right) \dsol,
\\
k_0 &=& \left[ \left( A - B \right) \epsilon \datm
+ \frac{1}{2} \left( A + B \right) \dsol \right]^2.
\ea
Since $k_0 > 0$,
the inequality~(\ref{in2}) does not tolerate $m_3 = 0$;
hence,
there is a non-trivial lower bound on the neutrino masses.
We want to find the numerical value of that bound.
Using the values of $A$ and $B$ in equations~(\ref{AB}),
one finds that
\ba
k_4 &=& - 16 c^2 s^2 \left( 1 - \cos{\chi} \right),
\\
k_2 &=& 2 \left( - 2 + 13 c^2 s^2 - 4 c^2 s^2 \cos{\chi} \right)
\epsilon \datm
\no & &
+ 3 \left( c^2 - s^2 \right)
\left( - 2 + 5 c^2 s^2 + 4 c^2 s^2 \cos{\chi} \right) \dsol,
\\
k_0 &=& \left[ 3 \left( c^2 - s^2 \right) \epsilon \datm
+ \frac{1}{2} \left( 5 - 10 c^2 s^2 - 8 c^2 s^2 \cos{\chi} \right)
\dsol \right]^2.
\ea
The case of real $x$,
$y$,
and $w$ corresponds to $\cos{\chi} = +1$.
In (and only in) that case,
the left-hand side of the inequality~(\ref{in2})
becomes linear in $m_3^2$;
besides,
in that case $k_2$ vanishes when $c^2 s^2 = 2 / 9$,
thereby generating singularities at the points $s^2 = 1/3$
and $s^2 = 2/3$,
as we saw in the previous section.

For $\cos{\chi} \neq +1$,
$k_4$ is negative.
Since $k_0$ is always positive,
the inequality~(\ref{in2}) then yields
\be
m_3^2 \ge \frac{\sqrt{k_2^2 + \left| k_4 \right| k_0} + k_2}
{\left| k_4 \right|} \equiv L.
\label{in3}
\ee
The task now consists in finding the minimum value of $L$
as a function of $\cos{\chi}$
(and of $\epsilon = \pm 1$);
that minimum value provides the lower bound on $m_3^2$.
It is easy to convince oneself that $L$
always has its minimum when $\cos{\chi} = -1$,
for all experimentally allowed values of $s^2$,
$\datm$,
and $\dsol$.
Computing $L$ as a function of $s^2$ for fixed $\cos{\chi} = -1$,
$\datm = 2.5 \times 10^{-3}~\mathrm{eV}$,
and $\dsol = 7.9 \times 10^{-5}~\mathrm{eV}$,
we conclude the following:
\begin{itemize}
\item When the neutrino mass spectrum is normal,
i.e.~when $\epsilon = +1$,
the minimum value of the lowest neutrino mass,
$m_1$,
hardly varies with $s^2$.
One has $m_1 > 1.679 \times 10^{-2}~\mathrm{eV}$ for $s^2 = 0.24$
and $m_1 > 1.665 \times 10^{-2}~\mathrm{eV}$ for $s^2 = 0.40$.
\item When the neutrino mass spectrum is inverted,
i.e.~when $\epsilon = -1$,
the minimum value of the lowest neutrino mass,
$m_3$,
varies strongly as a function of $s^2$.
One has $m_3 > 2.9 \times 10^{-2}~\mathrm{eV}$ for $s^2 = 0.24$,
$m_3 > 9 \times 10^{-3}~\mathrm{eV}$ for $s^2 = 0.40$.
\end{itemize}

Thus,
the mass matrix of equation~(\ref{mnu2}) with complex $x$,
$y$,
and $w$ is not very predictive:
it only allows one to derive a rather mild lower bound
on the neutrino masses.
There is also no prediction for the effective mass $m_{\beta \beta}$,
except for the rather trivial bounds
\be
\left| m_1^2 c^2 - m_2^2 s^2 \right|
\le m_{\beta \beta} \le
\left| m_1^2 c^2 + m_2^2 s^2 \right|.
\ee

\section{Conclusions}
\label{concl}

In this paper
we have constructed an extension of the Standard Model
with three Higgs doublets $\phi_\alpha$
and four scalar gauge triplets $\Delta_\alpha$ and $\Delta_4$.
The scalar triplets generate a type-II seesaw mechanism,
thus explaining the smallness of the neutrino masses.
We have employed a large horizontal symmetry group $G$,
generated by the permutation group $S_3$ of the indices $\alpha$
and by six cyclic groups $\mathbbm{Z}_2$.
After spontaneous symmetry breaking,
the charged-lepton mass matrix is diagonal;
the different VEVs of the $\phi_\alpha$
allow for different charged-lepton masses $m_\alpha$
($\alpha = e,\,\mu,\,\tau$).
In order to obtain a realistic neutrino mass matrix $\mnu$,
we additionally allow for soft breaking of $G$,
through terms of dimension two in the scalar potential.
A crucial feature of our model is the equality among
the diagonal entries of $\mnu$---this is
one of the reasons for the predictiveness of the model.

There are two relevant options:
breaking $G$ softly in the mass matrix of the scalar triplets
either fully or keeping a $\mu \leftrightarrow \tau$ symmetry intact;
and having either hard or spontaneous $CP$ breaking.
Our model has the interesting property that
spontaneous $CP$ violation has no effect on $\mnu$,
i.e.~it does not generate any physical phases in lepton mixing.
The most predictive scenario combines
the preservation of $\mu$--$\tau$ interchange symmetry
with spontaneous $CP$ violation,
in which case we arrive at a viable neutrino mass matrix
which has only three (real) parameters. 
This neutrino mass matrix leads to the usual predictions
of $\mu$--$\tau$ symmetric neutrino mass matrices,
namely maximal atmospheric mixing
and $\theta_{13}= 0$---hence no $CP$ violation
in neutrino oscillations. 
Besides,
the $CP$ property mentioned before
also prevents Majorana phases in our case.

The solar mixing angle $\theta$ is undetermined.
Our three-parameter neutrino mass matrix
predicts the neutrino masses $m_j$ 
as functions of the two mass-squared differences and of $\theta$. 
For $\sin^2 \theta < 1/3$, which seems to be preferred by the data,
we have a normal spectrum,
while for $\sin^2 \theta > 1/3$ the neutrino mass spectrum is inverted.
When $\sin^2 \theta \to 1/3$ all the $m_j$ diverge---see
figure~\ref{fig1}.
As for the effective mass $m_{\beta\beta}$ 
of neutrinoless $2\beta$ decay,
our three-parameter mass matrix predicts 
$m_{\beta\beta} \approx m_3/3$.

\paragraph{Acknowledgements:}
We thank Ernest Ma for suggestions and discussions
that contributed significantly to this work.
W.G.\ is grateful to H.~Urbantke for discussions on finite groups.
The work of L.L.\ was supported by the Portuguese
\textit{Funda\c c\~ao para a Ci\^encia e a Tecnologia}
through the projects POCTI/FNU/44409/2002,
POCTI/FP/63415/2005,
and U777--Plurianual.

\begin{appendix}

\section{The group structure of our model}
\label{group}
\setcounter{equation}{0}
\renewcommand{\theequation}{A\arabic{equation}}

In this appendix we attempt a mathematical description
of the full symmetry group of our model
and of its irreducible representations (irreps). 
Clearly,
$S_3$ commutes neither with the 
$\mathbf{z}_\alpha^{(1)}$ of equation~(\ref{z1}) 
nor with the 
$\mathbf{z}_\alpha^{(2)}$ of equation~(\ref{z2}),
thus the full symmetry group is rather complicated.

Let us define
\be
n_1 = \mathrm{diag} \left( -1,\, 1,\, 1 \right),
\quad
n_2 = \mathrm{diag} \left( 1,\, -1,\, 1 \right),
\quad
n_3 = \mathrm{diag} \left( 1,\, 1,\, -1 \right),
\ee
\be
c_+ = \left( \begin{array}{ccc}
0 & 0 & 1 \\ 1 & 0 & 0 \\ 0 & 1 & 0
\end{array} \right),
\quad
c_- = \left( \begin{array}{ccc}
0 & 1 & 0 \\ 0 & 0 & 1 \\ 1 & 0 & 0
\end{array} \right),
\ee
\be
t_1 = \left( \begin{array}{ccc}
1 & 0 & 0 \\ 0 & 0 & 1 \\ 0 & 1 & 0
\end{array} \right),
\quad
t_2 = \left( \begin{array}{ccc}
0 & 0 & 1 \\ 0 & 1 & 0 \\ 1 & 0 & 0
\end{array} \right),
\quad
t_3 = \left( \begin{array}{ccc}
0 & 1 & 0 \\ 1 & 0 & 0 \\ 0 & 0 & 1
\end{array} \right).
\ee
Then,
\newcommand{\bu}{\mathbbm{1}}
\be
N = \left\{ \bu,\, n_1,\, n_2,\, n_3,\, n_1 n_2,\, n_2 n_3,\, n_3 n_1,\,
- \bu \right\}
\ee
forms an Abelian group isomorphic to
$\mathbbm{Z}_2 \times \mathbbm{Z}_2 \times \mathbbm{Z}_2$.
Also,
\be
\hat S_3 = \left\{ \bu,\, c_+,\, c_-,\, t_1,\, t_2,\, t_3 \right\}
\ee
forms a three-dimensional (reducible) representation of $S_3$.

Let us call $G$ the symmetry group utilized in this paper.
$G$ may be defined to be the group of the $6 \times 6$ matrices
\be
\label{mns}
\left( \begin{array}{cc}
m s & 0 \\ 0 & n s
\end{array} \right),
\quad
m, n \in N,
\quad
s \in \hat S_3.
\ee
This defining \emph{reducible} representation of $G$
may be called the $\mathbf{6}$.
Clearly,
$G$ has $8 \times 8 \times 6 =  384$ elements.\footnote{The
$8 \times 6 = 48$ matrices $m s$,
where $m \in N$ and $s \in \hat S_3$,
form the Coxeter group $B_3$.
(We thank E.~Ma for drawing our attention to Coxeter groups.)
We may write $G = N \rtimes B_3 = \left( N \times N \right) \rtimes S_3$,
the symbol $\rtimes$ denoting a semi-direct product.}
Calling $\left( m, n, s \right)$ the abstract element of $G$
which is represented in the $\mathbf{6}$
by the matrix of equation~(\ref{mns}),
the group multiplication law is
\be
\label{law}
\left( m_1, n_1, s_1 \right) \left( m_2, n_2, s_2 \right)
= \left( m_1 s_1 m_2 s_1^{-1}, n_1 s_1 n_2 s_1^{-1}, s_1 s_2 \right).
\ee
From this group multiplication law
it follows that $G$ has eight one-dimensional irreps:
\be
\underline{1}^{(p,q,r)}: \quad
\left( m, n, s \right) \to
\left( \det{m} \right)^p \left( \det{n} \right)^q  \left( \det{s} \right)^r,
\quad \mbox{with} \quad p, q, r \in \left\{ 0, 1 \right\}.
\ee
It is obvious from equation~(\ref{mns})
that the matrices $m s$ give a three-dimensional irrep of $G$,
and similarly with the matrices $n s$.
The matrices $m n s$ give one further three-dimensional irrep of $G$,
since
\be
m_1 n_1 s_1 m_2 n_2 s_2 = m_3 n_3 s_3,
\quad \mbox{with} \quad m_3 = m_1 s_1 m_2 s_1^{-1}, \
n_3 = n_1 s_1 n_2 s_1^{-1}, \ \mbox{and} \ s_3 = s_1 s_2
\ee
complies with the multiplication law~(\ref{law}).
Thus,
$G$ has 24 three-dimensional irreps:
\ba
\underline{3}_1^{(p,q,r)}: & &
\left( m, n, s \right) \to 
\left( \det{m} \right)^p \left( \det{n} \right)^q \left( \det{s} \right)^r
m s, \\
\underline{3}_2^{(p,q,r)}: & &
\left( m, n, s \right) \to 
\left( \det{m} \right)^p \left( \det{n} \right)^q \left( \det{s} \right)^r
n s, \\
\underline{3}_3^{(p,q,r)}: & &
\left( m, n, s \right) \to 
\left( \det{m} \right)^p \left( \det{n} \right)^q \left( \det{s} \right)^r
m n s,
\ea
with $p, q, r \in \left\{ 0, 1 \right\}$.

The three-dimensional representations of $G$
that we employ in our model are
\be
\begin{array}{rcl}
m s & \mathrm{for} & \left( \phi_e, \,\phi_\mu,\, \phi_\tau \right), \\
m n s & \mathrm{for} & \left( e_R, \,\mu_R, \,\tau_R \right), \\
n s & \mathrm{for} & \left( D_{Le}, \,D_{L\mu}, \, D_{L\tau} \right), \\
\left( \det{n} \right) n s & \mathrm{for} &
\left( \Delta_e,\, \Delta_\mu,\, \Delta_\tau \right).
\end{array}
\ee

Our group $G$ also has two and six-dimensional irreps,
which are not used in our model.
Next we include, for completeness, their construction.

The group $S_3$ has a two-dimensional irrep $D_2$,
generated by
\be
t_1 \to \left( \begin{array}{cc}
0 & \omega \\ \omega^2 & 0 \end{array} \right),
\quad
t_2 \to \left( \begin{array}{cc}
0 & \omega^2 \\ \omega & 0 \end{array} \right),
\ee
where $\omega = \left( -1 + i \sqrt{3} \right) /\, 2$.
Note that $\left( \det{s} \right) D_2 \left( s \right)$
is isomorphic to $D_2 \left( s \right)$:
\be
\left( \det{s} \right) D_2 \left( s \right) =
\left( \begin{array}{cc} 1 & 0 \\ 0 & -1 \end{array} \right)
D_2 \left( s \right)
\left( \begin{array}{cc} 1 & 0 \\ 0 & -1 \end{array} \right).
\ee
Therefore,
$G$  has four two-dimensional irreps:
\be
\underline{2}^{(p,q)}: \quad \left( m, n, s \right) \to 
\left( \det{m} \right)^p \left( \det{n} \right)^q D_2 \left( s \right),
\quad \mbox{with} \quad p,q \in \left\{ 0, 1 \right\}.
\ee

The remaining irreps of $G$ are four six-dimensional ones,
\be
\underline{6}^{(p,q)}: \quad \left( m, n, s \right) \to
\left( \det{m} \right)^p \left( \det{n} \right)^q  D_6 \left( m, n, s \right),
\quad \mbox{with} \quad p,q \in \left\{ 0, 1 \right\}.
\ee
The irrep $D_6 \left( m, n, s \right)$
is found in the decomposition
of the product of the irreps $m s$ and $n s$.
Suppose there is a space $\mathbbm{C}^3$
spanned by $e_{1,2,3}$ transforming like $m s$,
and another space $\mathbbm{C}^3$
spanned by $e^\prime_{1,2,3}$ transforming like $n s$.
Then,
the space spanned by the $e_k \otimes e^\prime_k$ 
($k = 1, 2, 3$)
transforms like $m n s$,
while the $e_j \otimes e^\prime_k$ with $j \neq k$ span a space
which transforms like $D_6 \left( m, n, s \right)$.
It can be shown that this representation
$D_6 \left( m, n, s \right)$ of $G$ is irreducible,
and also that it is equivalent to
$\left( \det{s} \right) D_6 \left( m, n, s \right)$.

The group $G$ has the interesting property
that it has no faithful irreps.
It is obvious that the irreps
with dimensions three and lower are not faithful.
The six-dimensional irreps are not faithful either,
as we now explain.
Defining the elements $a$ and $b$ of $G$ by 
$a = \left( -\mathbbm{1}, \mathbbm{1}, \mathbbm{1} \right)$
and $b = \left( \mathbbm{1}, -\mathbbm{1}, \mathbbm{1} \right)$,
then $a$,
$b$,
and $ab$ generate the subgroups $\mathbbm{Z}_2^{(a)}$,
$\mathbbm{Z}_2^{(b)}$,
and $\mathbbm{Z}_2^{(ab)}$ of $G$,
respectively.
The isomorphisms
\be
\underline{6}^{(0,0)} \cong G\,/\,\mathbbm{Z}_2^{(ab)}, \quad
\underline{6}^{(1,0)} \cong G\,/\,\mathbbm{Z}_2^{(a)}, \quad
\underline{6}^{(0,1)} \cong G\,/\,\mathbbm{Z}_2^{(b)}, \quad
\underline{6}^{(1,1)} \cong G\,/
\left( \mathbbm{Z}_2^{(a)} \times \mathbbm{Z}_2^{(b)} \right)
\ee
are easy to demonstrate.
Thus,
none of the six-dimensional irreps represents $G$ faithfully.

\section{Spontaneous breaking of the $\mu$--$\tau$ symmetry}
\label{SSB}
\setcounter{equation}{0}
\renewcommand{\theequation}{B\arabic{equation}}

Let us consider a simplified model with only two VEVs,
$v_\mu$ and $v_\tau$.
We assume the following symmetries: 
\be
\begin{array}{cl}
z_1: & v_\mu \to - v_\mu, \ v_\tau \to v_\tau; \\
z_2: & v_\mu \to v_\mu, \ v_\tau \to - v_\tau; \\
z_3: & v_\mu \leftrightarrow v_\tau.
\end{array}
\ee
The symmetries $z_{1,2}$ are assumed to be softly broken
by terms of dimension two,
while $z_3$ is assumed to be exactly conserved.
For the sake of clarity
we also assume all coefficients to be real.
Then,
\be
V_0 =
a \left( \left| v_\mu \right|^2 + \left| v_\tau \right|^2 \right)
+ b \left( v_\mu^\ast v_\tau + v_\tau^\ast v_\mu \right)
+ \lambda \left( \left| v_\mu \right|^2
+ \left| v_\tau \right|^2 \right)^2
+ \lambda^\prime \left| v_\mu \right|^2 \left| v_\tau \right|^2. 
\ee
Only the $b$ term breaks $z_{1,2}$ softly.

Without loss of generality we take $v_\mu$ to be real and positive,
writing
\be
v_\mu = \nu \cos{\phi}, \quad v_\tau = \nu \sin{\phi}\, e^{i\alpha},
\ee
with $\nu > 0$ and $\phi$ in the first quadrant.
Then we obtain
\ba
V_0 &=& a \nu^2 + b \nu^2 \sin{2 \phi} \cos{\alpha} + \lambda \nu^4
+ \frac{\lambda^\prime \nu^4}{4}\, \sin^2{2 \phi}.
\label{vvv}
\ea
We require that
\be
0 = \frac{\partial V_0}{\partial \left( 2 \phi \right)} =
\left( \nu^2 \cos{2 \phi} \right) \left( b \cos{\alpha}
+ \frac{\lambda^\prime \nu^2}{2}\, \sin{2 \phi} \right).
\ee
The solution $\cos{2 \phi} = 0$ corresponds to
$\left| v_\mu \right| = \left| v_\tau \right|$ and is undesirable.
But there is another solution,
\be
\sin{2 \phi} = - \frac{2 b \cos{\alpha}}{\lambda^\prime \nu^2},
\ee
which we adopt.
Since the minimization of $V_0$ in equation~(\ref{vvv})
with respect to $\alpha$
leads to $b \cos{\alpha} = - \left| b \right|$ being negative,
we must assume $\lambda^\prime$ to be positive.
If
\be
\left| b \right| \ll \frac{\lambda^\prime \nu^2}{2},
\ee
which corresponds to the soft-breaking term
being very small, then $\sin{2 \phi} \ll 1$
and $\left| v_\mu \right| \ll \left| v_\tau \right|$ 
can be realized.

\end{appendix}

\end{document}